\documentclass[sigconf]{acmart}

\renewcommand\footnotetextcopyrightpermission[1]{} 
\usepackage{graphicx}
\usepackage{listings}
\usepackage{url}
\usepackage{subfigure}
\usepackage{algorithm2e}
\usepackage{amsmath}
\usepackage[T1]{fontenc}
\usepackage{appendix}
\usepackage{footnote}
\makesavenoteenv{tabular}
\makesavenoteenv{table}

\AtBeginDocument{%
  \providecommand\BibTeX{{%
    \normalfont B\kern-0.5em{\scshape i\kern-0.25em b}\kern-0.8em\TeX}}}

\setcopyright{none}

\begin{document}
%
\title{Generating Comprehensive Data with Protocol Fuzzing for Applying Deep Learning to Detect Network Attacks}
\author{Qingtian Zou}
\email{qzz32@csus.edu}
\affiliation{%
  \institution{The Pennsylvania State University}
}

\author{Anoop Singhal}
\email{anoop.singhal@nist.gov}
\affiliation{%
  \institution{National Institute of Standards and Technology}
}

\author{Xiaoyan Sun}
\email{xiaoyan.sun@csus.edu}
\affiliation{%
  \institution{California State University, Sacramento}
}

\author{Peng Liu}
\email{pxl20@csus.edu}
\affiliation{%
  \institution{The Pennsylvania State University}
}

\begin{abstract}
Network attacks have become a major security concern for organizations worldwide and have also drawn attention in the academics. Recently, researchers have applied neural networks to detect network attacks with network logs. However, public network data sets have major drawbacks such as limited data sample variations and unbalanced data with respect to malicious and benign samples. In this paper, we present a new approach, protocol fuzzing, to automatically generate high-quality network data, on which deep learning models can be trained. Our findings show that fuzzing generates data samples that cover real-world data and deep learning models trained with fuzzed data can successfully detect real network attacks.
\end{abstract}


\maketitle

\section{Introduction}
\label{sec:intro}
As one of the most prevalent ways to compromise enterprise networks, network attack remains a prominent security concern. It can lead to serious consequences such as large-scale data breaches, system infection, and integrity degradation, particularly when network attacks are employed in attack strategies such as advanced persistent threats (APT). Therefore, detecting network attacks early on is important to prevent potential negative impacts towards the enterprise networks. 

The network attack detection methods can be classified into two categories: {\em host-independent} methods and {\em host-dependent} methods. The host-independent methods solely rely on the network traffic, while the host-dependent methods \cite{Hogan_Johnson_Halappanavar_2013,Kumar_Tapaswi_2012,Arote_Arya_2015,Sun_Chang_Chang_Lin_2009,Yuan_Kant_Mohapatra_Chuah_2006,Goswami_Hoque_Bhattacharyya_Kalita_2017} depend on additional data collected on the victim hosts. An example of host-dependent method is detecting pass the hash (PtH) by monitoring the account's activities~\cite{iainfoulds2020Sep} on the victim hosts, as recommended by Microsoft. The host-dependent methods have some evident drawbacks: they have fairly high deployment costs and operation costs; they are also error-prone due to necessary manual configuration by human administrators. Therefore, the host-independent detection methods are highly desired as they can significantly reduce deployment and operation costs while letting the detection system have a smaller attack surface.

However, we found that the existing host-independent methods, including the classical intrusion detection approaches, often fall short of detecting some well-known and commonly used 
network attacks, such as the PtH attack, ARP poisoning attacks, DNS cache poisoning attacks 
and TELNET session hijacking attacks. The 
signature-based~\cite{Kaur_Singh_2013,Taylor_Harrison_Krings_Hanebutte_McQueen_2001} methods could be easily 
evaded by slightly changing the attack payloads. For some network attacks, it's not even 
possible to find any signatures. The anomaly-detection-based~\cite{Amini_Jalili_Shahriari_2006} methods tend 
to raise lots of false positives, which prevent them from being practically deployed in real world. 


Given the evident drawbacks of host-dependent methods and the vital role played 
by network attacks in APT, it becomes increasingly important to develop 
highly accurate {\bf host-independent methods} for detecting important network attacks 
(e.g., the PtH attack, DNS cache poisoning attacks, and session hijacking attacks)
that do not have a distinct signature. 
For this purpose, researchers have started applying deep learning to train host-independent 
neural network detection models. 
However, a fundamental challenge is that neural networks usually require quality network 
traffic data and correct labels, which are hard to obtain in real world.
First, training neural networks for classification needs labeled data. However, in real world, network traffics are flooded with benign packets, which makes it difficult to label malicious network packets. Since lack of ground-truth is a challenge widely recognized in the network security community, how to assign a correct label to each packet in real-world network traffic is still an open problem.  
Second, public data sets~\cite{BibEntry2020Jan,dhanabal2015study,Moustafa2015,pfahringer2000winning,Sharafaldin2018} for network attacks are barely useful. Most data sets are generated (and synthesized) with various simulated or emulated types of benign and attack activities, and each type of attack is only launched for a few times. Therefore, these data sets are highly unbalanced and have limited variations, which downgrades the trained neural networks. In a word, it is very challenging to get high-quality labeled data sets on which neural networks can be trained to detect network attacks.

In this paper, we propose a new approach, protocol fuzzing, to address the data challenges mentioned above, so that the deep learning models can be trained for network attack detection. For a client and a server program communicating with a specific network protocol, protocol fuzzing means the client generating mutated network packets as input to be fed into the server program. With protocol fuzzing, a large variety of {\em malicious} network packets for a chosen network attack can be generated at a fast speed. Since the network packets are all generated from the chosen network attack, they can be labeled automatically without much human efforts. Protocol fuzzing can generate data with more variations than real world data, or even data that are not yet observed in real world. Moreover, these merits remain when protocol fuzzing is leveraged to generate the needed {\em benign} network packets. It should be noted that our method is different from data synthesis. Data synthesis is to enhance existing data~\cite{jan2020throwing}, while our method is to generate new data. The new approach enables generating high-quality network traffic data with less time compared to the traditional data generation methods.

The main contributions of this work are as follows: 1) We propose a new approach for automatically generating high-quality data for applying deep learning to detect network attacks. 2) By proposing the notion of \emph{covered by}, we seek to take the first step towards systematically assessing the comprehensiveness of the training data used to detect network attacks. 3) We have trained and evaluated {\bf host-independent} deep learning models with our generated data and verified that these models can detect a set of network attacks with high accuracy.  


\vspace{-1mm}
\section{Related Work}
\label{sec:related_works}
The research community has been tackling the network attack detection problem with both classical and novel approaches. In this section, we will discuss the research works related to network attack detection from different perspectives.

{\bf Traditional network attack detection approaches.}
Traditionally, people usually detect network attacks with approaches like signature-based, rule-based, and anomaly detection-based methods, some of which are still in use today. In the past, signature-based intrusion detection system (IDS) usually manually crafted signatures~\cite{Taylor_Harrison_Krings_Hanebutte_McQueen_2001}, which heavily depends on manual efforts. Nowadays, people focus more on automatically generate signatures~\cite{Kaur_Singh_2013}. However, signatures need to be constantly updated to align with newer attacks and signature-based detections can be easily evaded by slightly changing the attack payload. Similar problems also exist for rule-based methods~\cite{Choi_Choi_Ko_Kim_2014}, which constantly need updates to the rules. As for anomaly detection-based methods, though they require much less manual efforts for updating, they tend to raise more false positives~\cite{Amini_Jalili_Shahriari_2006}, which interrupt organizations' normal operations.

{\bf Deep learning for network attack detection.}
Network attacks are essential for APTs. Some common network attack types include probing, DoS, Remote-to-local (R2L), etc. The traditional methods for network attack detection include signature-based, rule-based, and anomaly detection. However, signature-based methods can be easily fooled by slightly changing the attack payload; rule-based methods need experts' knowledge to regularly update rules; and anomaly detection methods suffer from high false positives. Therefore, the deep learning methods are recently adopted for network attack detection. Some deep learning based research works focus on just one type of network attack and do binary classifications. An example is DeepDefense~\cite{Yuan2017}, which employs deep learning to detect distributed DoS (DDoS) attacks. Others~\cite{Faker2019,Millar2018,Yin2017,Zhang2019} try multi-class classifications, which include one benign class and multiple malicious classes for different kinds of network attacks. The above mentioned research works all use public data sets. 

{\bf Network data sets for training and testing detection models.} 
To apply deep learning for network attack detection, a data set is required on which the model can be trained. Commonly used public data sets include KDD99~\cite{pfahringer2000winning}, NSL-KDD~\cite{dhanabal2015study}, UNSW-NB15~\cite{Moustafa2015}, CICIDS2017~\cite{Sharafaldin2018}, and CSE-CIC-IDS2018~\cite{BibEntry2020Jan}. The public data sets are all generated in test-bed environments, with simulated benign and malicious activities. The data set is often unbalanced due to overwhelming amount of benign data. Even for only malicious activities, multiple types of attacks may be included and the amount of malicious data for each attack type varies a lot. Unbalanced data samples can pose problems for neural networks because they may lead to biased trained models. Besides, it is very difficult to label the unbalanced data. There are approaches to mitigate this kind of unbalancing, such as downsampling and oversampling. However, downsampling mitigates the unbalancing by reducing the number of data samples, and oversampling does it by introducing duplicate data samples, so it is better to start with a balanced data set.

{\bf Protocol fuzzing.}
Fuzzing is originally a black-box software testing technique, which looks for implementation bugs by feeding mutated data. A key function of fuzzers is to generate randomized data which still follows the original semantics. There are tools for building flexible and security-oriented network protocol fuzzers, such as SNOOZE~\cite{Banks2006}. Network protocol fuzzing frameworks such as AutoFuzz~\cite{Aitel2002,Gorbunov2010} have also been presented. They either act as the client, constructing packets from the beginning, or act as a proxy, modifying packets on the fly.
In this paper, we use protocol fuzzing for a different purpose, which is to directly generate high-quality data sets that can be used to train neural networks. With this purpose in mind, instead of using the tools or frameworks listed above, we decide to prepare our own scripts based on the scripts for normal behaviors or attacks, so that the fuzzing can be applied as direct as possible. Typical fuzzing hopes to trigger program bugs, but we hope to generate various and a large amount of data while keeping the benign/malicious activities successful.

\vspace{-1mm}
\section{Problem Formulation}
\label{sec:NA_selection}

\begin{figure}
    \centering
    \includegraphics[width=0.42\textwidth]{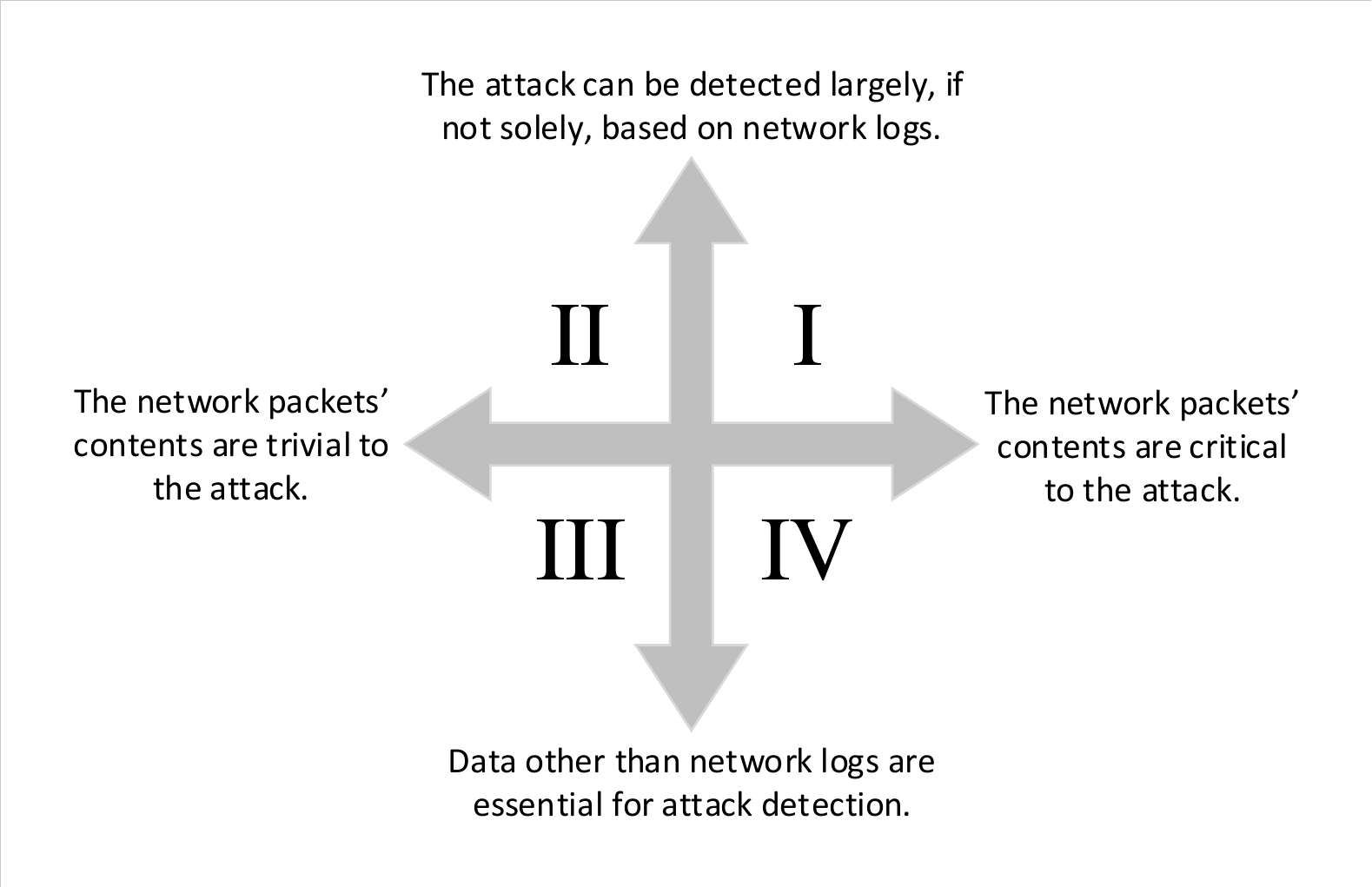}
        \vspace{-4mm}
    \caption{Network attack categories.}
    \label{fig:na-categories}
    \vspace{-7mm}
\end{figure}

In principle, whether deep neural networks can play an essential role depends on the characteristics of the particular network attack in concern. To provide a more general guideline, 
we categorize network attacks into four categories like a Cartesian coordinate system as presented in Figure~\ref{fig:na-categories}. All network attacks are categorized into four quadrants, based on two criteria: (a) whether the network packets' contents are critical to the attack; (b) whether network logs are the key for network attack detection and whether they are sufficient, which means, whether some data other than network logs actually plays a critical role for the detection.

We argue that the network attacks that fall into Region I are most likely to benefit from a carefully trained deep neural network. Accordingly, 
the protocol fuzzing for data generation is most useful for detecting the network attacks with the following characteristics: 1) The attack can be detected largely, if not solely, based on network logs. 2) The network packets' contents are critical to the attack. 3) There is not a distinctive signature for detecting the attack.

The rationale behind is as follows: (a) Protocol fuzzing is to generate fuzzed network traffic, which is recorded in the network logs. If using solely network traffic is not sufficient and some other data is essential to detect the network attack, then protocol fuzzing has limited contribution to the attack detection. (b) Protocol fuzzing is to fuzz the network packets' contents, which means mutating the data in the packets. If the packets' contents are not important, then again fuzzing does help much with the attack detection. For example, if an attacker launches lateral movement attacks with stolen account username and password, the attack detection is more about monitoring the abuse of accounts. Network logs are not helpful in this case because the attacker is using legitimate credentials and valid tools. Another example is resource exhaustion DoS attacks. The network logs can be used to detect such attacks, but the packets' contents are of little use. The attack can be achieved by sending huge amounts of packets in a short time period, without the need of significant changes in packet contents. 


Therefore, this paper will answer the question: is Protocol Fuzzing an effective approach in generating comprehensive data and trustworthy labels for applying deep learning to detect Region I attacks?  

Without loss of generality, we have selected four representative network attacks in Region I, which are PtH attacks, address resolution protocol (ARP) poisoning attacks, domain name system (DNS) cache poisoning attacks, and TELNET session hijacking attacks. \textbf{PtH} is a well-known attack for lateral movements, and has been reported to be used in more than ten APT campaigns in the recent years~\cite{BibEntry2020Jun}. \textbf{ARP poisoning attacks} and \textbf{DNS cache poisoning attacks} are difficult to be identified by traditional approaches because both attacks involve spoofing. That is, attackers intentionally make the packets seem to be benign. Though TELNET is rarely used nowadays, \textbf{TELNET session hijacking} is one of the representatives for Transmission Control Protocol (TCP) session hijacking attacks, which are still commonly used today.

The four attacks can be detected by traditional approaches to some extent. For example, PtH may be detected by monitoring user login and login methods; ARP poisoning may be detected by monitoring the ARP table. However, such approaches are only effective after the attacks have succeeded, and they need to access certain log data on the victim machine. If the attacks can be detected at the network level, we can deploy the detection model as an NIDS (network IDS) in the enterprise LAN and proactively raise alarms. Other tools like Snort~\cite{Snort} may be able to detection those network attacks at the network level, but they are hardly effective against unseen attacks even of the same category, and their rules or signatures need to be constantly updated. That is why neural networks can be used as an alternative. To demonstrate the effectiveness of our approach, we have applied protocol fuzzing to generate network logs, trained and evaluated neural networks to detect the four attacks respectively based on the collected network logs. 



\vspace{-3mm}
\section{Generating Comprehensive Data}
\label{sec:data-gen}

In this section, we will discuss technical details about how to generate benign and malicious data sets for the four network attacks, including PtH, ARP poisoning, DNS cache poisoning, and TCP session hijacking. We'll firstly introduce the general approach and implementation principles of protocol fuzzing, and then discuss the data generation for the four attacks respectively. All four attacks are carried out thousands of times so that a fair amount of malicious data can be collected. Benign data generation also lasts long enough to gather the commensurate amount of data compared to the malicious data. The network packet capturing is performed at the victim's side. 


\vspace{-4mm}
\subsection{Protocol Fuzzing and The Implementation}
\label{sec:data-gen-general}

In client-server enterprise computing, the server side protocol implementations are often complex and error-prone, and the clients are usually designed in accordance with the server. Hence, there is a need to achieve thorough testing of the server side implementation. That is why protocol fuzzing tools~\cite{Aitel2002,Banks2006,Gorbunov2010} are usually functioning at the client side, sending mutated network packets to the tested server programs, or acting as a proxy modifying packets on the fly and replaying them, so that unexpected errors on the tested server programs may be triggered. A main difference between protocol fuzzing and software fuzzing is that the protocol specification, especially its state transition diagram, will be used to guide the fuzzing process. In this way, fuzzing tests could be performed in a stateful manner. 


In this paper, we leverage protocol fuzzing to change the contents of network packets, specifically, the values of some fields in the packets. If we are to fuzz a field, we will assign values chosen by ourselves, instead of the values chosen by the network client program. The fields to be fuzzed are chosen based on the following steps, as shown in Algorithm~\ref{alg}: 1) We list the fields in the packet of the attack-specific protocol. 2) We pick one field on the list and fuzz it by assigning values of our choices, rather than values that are normally provided by the network programs. 3) We monitor how the attack goes after fuzzing the field. If the attack success rate is above 50\%, we confirm that this field can be fuzzed. 4) After one field is fuzzed, we move on to the next field on the list, while keeping the already fuzzed field(s) still fuzzed. As for how the values are assigned (how the fields are fuzzed), details will be provided in the later paragraphs. 

\begin{algorithm}
    \SetAlgoLined
    \KwResult{\texttt{BList}, which stores fields to fuzz}
     input \texttt{AList} of all available fields\;
     initialize an empty \texttt{BList} to store fields to fuzz\;
     \ForEach{\texttt{field} in \texttt{AList}}{
         fuzz \texttt{field}\;
         fuzz all fields in \texttt{BList}\;
         launch the attack for hundreds of times\;
         count successful attacks and calculate success rate\;
         \If{attack success rate is over 50\%}{
            add \texttt{field} to \texttt{BList}\;
             }
     }
     \caption{Select fields to be fuzzed.}
\label{alg}
\end{algorithm}

Note that we have double insurance to keep the fuzzed packets valid. The first insurance is before we choose the fields to be fuzzed, we make sure the $AList$ does not contain fields that will affect the packets' basic integrity, such as fields of checksum values and packet lengths. The values of those fields are determined only after other fields' values are all determined, so that the packets' integrity is preserved. The second insurance is when we choose the fields to be fuzzed, we make sure the network attack success rate is always above 50\%, so that the sessions' integrity is preserved.

We implemented a number of Python scripts to fuzz most fields and send the fuzzed network packets with open-source libraries such as scapy. The fields are fuzzed by assigning values randomly in their valid ranges. We will first take a look of the unfuzzed packet structure and corresponding documents, learn the valid range of that field, and then assign random values within that range. For example, one of the fields that are often fuzzed is the \texttt{time to live} value in the \texttt{IP} layer. By looking at the genuine packets and related documents, we learn that it is an 8-bit field, and the valid values are integers within $[0,255]$. Hence, a random integer is generated from the range $[0,255]$ and assigned to that field when fuzzing it. If the field to fuzz is a binary flag field, then the valid values only include 0 and 1, so we randomly generate an integer from range $[0,1]$ and assign it to that field. Sometimes, binary flags are adjacent bits in the network packet, such as the \texttt{flags} field in the \texttt{IP} layer. For these, we treat all adjacent bits as one field and assign values to the entire field, instead of performing fuzzing to each individual bits respectively. Take the \texttt{flags} field as an example. It is a 3-bit field, and the first bit is reserved and must be 0, so the valid values are ${0,2,4,6}$. When fuzzing those flags, we assign the whole 3-bit field with a randomly chosen value from the valid range, rather than assigning values to the last two bits separately. 

However, some fields, such as the MAC addresses and IP addresses, may not be fuzzed as mentioned above using the Python scripts. 1) One reason is that some fields are not accessible within the Python scripts due to the limitation of the scapy library we use. The fields in the \texttt{ETH} layer cannot be accessed. For those fields, we have created some additional scripts to complete the fuzzing. For example, when fuzzing MAC address, we will have another Bash script to randomly generate a MAC address, shut down the target network interface, re-initialize the network interface with the randomized MAC address, and then bring up the network interface. This Bash script is configured to auto-run periodically with a given time interval, so that the MAC address can be fuzzed in different network sessions. 2) Another reason not to use python scripts for some fields is that special procedures may be needed to assign valid values to these fields. For example, the IP addresses are not fuzzed using the Python scripts, even though they are accessible in the Python scripts. The IP addresses cannot be fuzzed by simply choosing a value randomly. Instead, a procedure is needed in order to assign valid values to the IP address field. Therefore, we deploy a Dynamic Host Configuration Protocol (DHCP) server in the local area network (LAN) to distribute IP addresses to machines connected to the LAN. If we are to fuzz the IP address of a host, we will connect it to the LAN, and make the DHCP server restart at a given time interval. Meantime, we'll flush the DHCP records with Bash scripts, so that the host will need to communicate with the DHCP server often for a new IP address.

Though the fields chosen to be fuzzed and the assigned values may be intuitively irrelevant to the success of attack, these modifications to the packets can make data samples misclassified as benign, which is the goal of the attacker. That is, to achieve network attacks without being detected. However, by introducing protocol fuzzing generated network packets, these overlooked data samples can be covered in the data set. 


\vspace{-4mm}
\subsection{PtH Data Generation}
\label{subsec:data-gen_PtH}

\begin{figure*}[ht]
\vspace{-2mm}
    \centering
    \includegraphics[width=\textwidth]{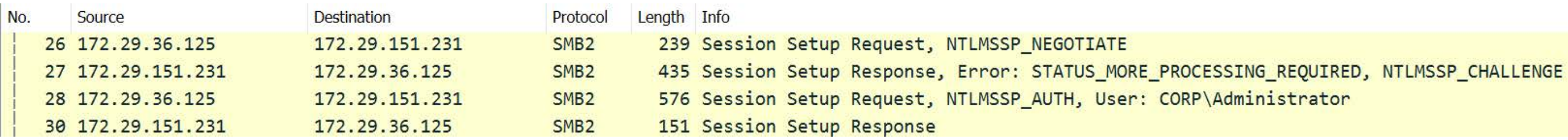}
    \vspace{-6mm}
    \caption{A subset of network packets during PtH attack.}
    \vspace{-6mm}
    \label{fig:authentication}
\end{figure*}

\begin{figure*}
    \centering
    \includegraphics[width=\textwidth]{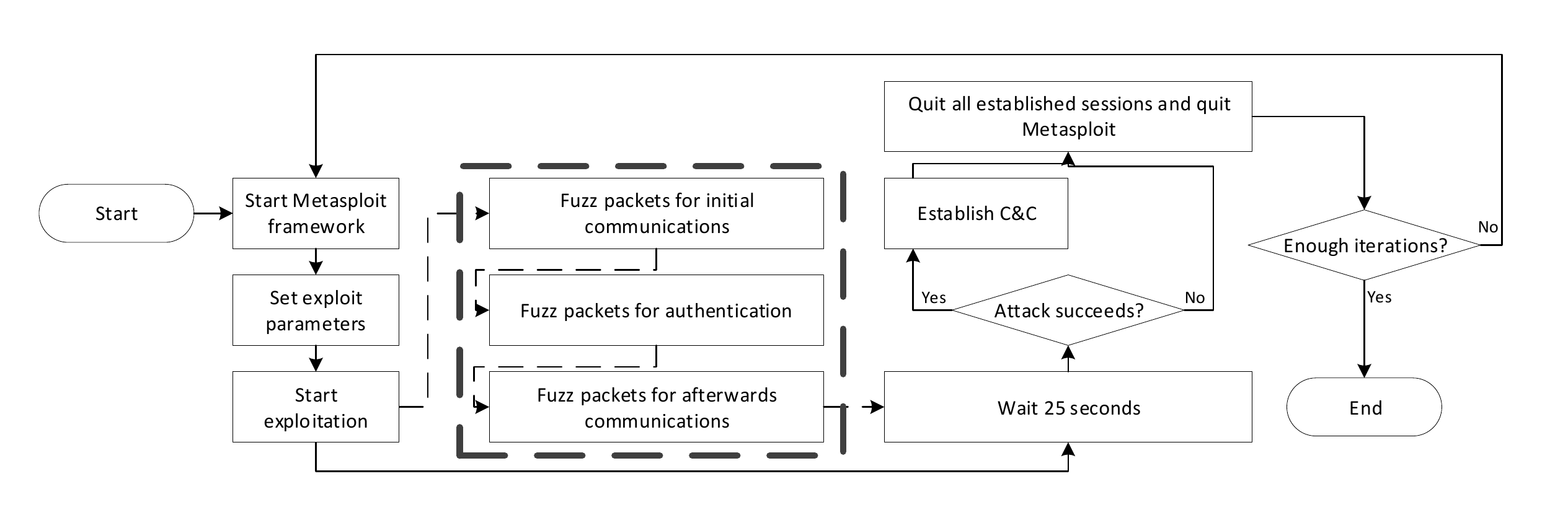}
     \vspace{-11mm}
    \caption{Fuzzing process for generating malicious PtH network data.}
    \vspace{-3mm}
    \label{fig:pth-mal-fuzz-proc}

\end{figure*}

PtH is a well-known technique for lateral movement. In remote login, plain text passwords are usually converted to hashes for authentication. Some authentication mechanisms only check whether hashes or the calculation results of them matches or not. PtH relies on these vulnerable mechanisms to impersonate a normal user with dumped hashes. In a client server model, we assume that: (a) normal users use benign client programs that are usually authenticated through mechanisms other than just using hashes, and that (b) attackers cannot get the plain text passwords and have to rely on hashes to impersonate a normal user. We can capture the network packets at the server side and find out which kind of authentication mechanism is used. Login sessions that use those vulnerable authentication mechanisms can then be identified as PtH attack.

Windows remote login processes, if not properly configured, can use such vulnerable authentication mechanisms. We set up a Windows 2012 Server R2 as the victim server machine, a Windows 7 as the user client machine, and another Kali Linux as the attacker machine. Windows remote login can be divided into three stages, protocol and mechanism negotiation (initial communication), authentication, and task-specific communication (afterwards communication). Each stage contains multiple network packets, and hashes are used in the authentication stage for impersonation. The authentication stage can be viewed as a sequence made up of client's authentication request, server's challenge, client's challenge response and server's authentication response, as shown in Figure~\ref{fig:authentication}. The client first sends a session setup request to the server; then the server responds to the client with a challenge; on receiving the challenge, the client uses the challenge and hashes to do calculations and sends back the result in challenge response packet; finally, the server verifies the result and sends back authentication response indicating whether authentication succeeds or not.

The data sets are automatically generated by protocol fuzzing. More than 15 fields are fuzzed in each SMB/SMB2 packets, including \texttt{SMB flags}, \texttt{SMB capabilities}, and fields in SMB header, etc. During a PtH attack, before a packet is sent to the server, we fuzz those fields in the application layer (\texttt{SMB/SMB2} layer for PtH attack) of the network packets. In this way, (a) the packet structure remains intact, so that the server will not discard the packet; (b) the authentication of PtH may be affected; (c) we can get a variety of network packets, and possibly, a variety of network packet sequences so the data diversity is improved.

Figure~\ref{fig:pth-mal-fuzz-proc} shows the fuzzing process for generating malicious data. We leverage the PtH script in Metasploit Framework to launch the attack. Boxes connected with solid lines are what happens at foreground, and boxes in the dash line area happen behind the scene. The process is to start the Metasploit Framework, set exploit parameters, start the exploitation, and then wait 25 seconds while monitoring the attack status. Waiting is necessary because the exploitation needs some time to complete. If the waiting time is too short, the attack may be stopped before completion. While the console is waiting at the foreground, the exploitation is on going at the background with fuzzed network packets. Network packets in all the three stages, initial communication, authentication, and afterwards communication, are fuzzed. After the exploitation, based on whether the attack succeeds or not, we may continue to establish C\&C, like what a real attacker will do. (The C\&C network traffic are mainly TCP packets, which are not used for attack detection. Details are discussed in subsection~\ref{subsec:pth-nn}.) Finally, we quit all possibly established sessions and the Metasploit Framework, and then either start another fuzzing iteration to generate more data or stop. The sign of a successful PtH attack is an established reverse shell, which can be observed at the attacker's side.

The same fuzzing method has also been applied in the generation of benign data from normal network traffic. For the benign scenario, we first prepare a list of normal commands, including files reading, writing, network interactions, etc. For each benign fuzzing iteration, we first randomly choose a command from the list, and then use valid username, plain-text password, and tool to log in to the server and execute the command.

All the network packets from malicious and benign network traffic are captured using Wireshark at the victim server's side. Due to fuzzing, not all PtH attempts or benign access attempts  can be guaranteed to be successful. For failed PtH attempts, we remove them from malicious data because they do not generate real malicious impact. These packets cannot be categorized as benign either because they are generated with attacker tools for malicious purpose, rather than for legitimate access. For failed benign accesses, we keep them in benign data, because normal user can also have failed logins due to typos, wrong passwords, etc.

\vspace{-3mm}
\subsection{ARP Poisoning Data Generation}
\label{subsec:data-gen_ARP}
When an Internet Protocol (IP) datagram is sent from one host to another in a LAN, the destination IP address must be resolved to a MAC address for transmission via the \texttt{ETH} layer. To resolve the IP address, a broadcast packet, known as an ARP request, is sent out on the LAN to ask which MAC address matches the destination IP. The destination host possessing this IP will respond with its MAC address in the ARP reply. ARP is a stateless protocol. Machines usually automatically cache all ARP replies they receive, regardless of whether they have requested them or not. Even ARP entries that have not yet expired will be overwritten when a new ARP reply packet is received. The host cannot verify where the ARP packets come from. This behavior is the vulnerability that allows ARP poisoning to occur. By spoofing ARP responses, the attacker can trick the victim into falsified mappings between IP addresses and MAC addresses, and thus intervening the network communication. For example, a machine within a LAN usually sends packets to the outside network through a router. By ARP poisoning, the attacker in the LAN may be able to redirect all traffic from and to the user machine to its own attacker machine. For example, at the user side, the attacker may map the router's IP address with the attacker's MAC address so that the traffic sent to the router is redirected to attacker. Similarly, at the router side, the user machine's IP address is mapped with the attacker's MAC address so that the traffic sent to the user machine is also redirected to the attacker. The attacker only needs to forward packets from the user machine and router to their intended targets, so that the attacker capture all network traffic without interrupting the network communication.


For this attack, our testbed constructs a LAN that contains an Ubuntu machine configured as a router with DHCP server enabled, another Ubuntu as the user machine, and a Kali Linux as the attacker machine. Fuzzed fields include \texttt{MAC address} and \texttt{IP address}. The attacker machine is configured to send out fuzzed ARP responses with randomly generated IP addresses and MAC addresses periodically. The MAC address and IP address fuzzing methods was discussed in Section~\ref{sec:data-gen-general}. The victims (``router'' and user machine) are configured to flush the IP-MAC mapping periodically. The IP addresses in spoofed packets are randomly generated from the same IP ranges distributed by the DHCP server, because the attacker is only interested in directing the traffic to an IP which actually exists in the LAN. On the user machine, we periodically run command $arp -a$ to print out the MAC-IP table. If MAC address in the table is not a valid machine in the LAN, then the attack is successful.

For benign data set generation, the user machine is configured to probe the presence of other machines by sending out ARP requests to all IP addresses within the LAN's IP range periodically so that enough data can be generated. In both the malicious and benign scenarios, the DHCP server is configured to flush the IP assignments to all machines within the LAN periodically, so that all machines' IP addresses within the LAN, except the DHCP server itself, will be changed in different iterations.

\vspace{-4mm}
\subsection{DNS Cache Poisoning Data Generation}
\label{subsec:data-gen_DNS}

A major functionality of Domain Name Service (DNS) is to provide the mapping between the domain names and IP addresses. When a client program on a user machine refers to a domain name, the domain name needs to be translated to an IP address for network communication. The DNS servers are responsible to perform such translation. 

The DNS system has a hierarchical structure that contains root name servers, top-level domain name servers, and authoritative name servers. Some examples are the public DNS servers \url{8.8.8.8} and \url{8.8.4.4} provided by Google, and recently released \url{1.1.1.1} by Cloudflare. These name servers, referred as the global DNS servers, provide records that maps the domain names and IP addresses. Due to the geological distance between user machines and the global DNS servers, it is very costly to contact the global DNS servers every time when mapping is needed. To reduce the cost, organizations often deploy their own DNS servers, referred as local DNS servers, within the LAN to cache the most commonly used mappings between domain names and IP addresses. Generally, when a user machine needs to make connection with a destination machine, it will contact the local DNS server first to resolve the domain name. If the local DNS server does not cache the DNS record for this domain name, it will send out a DNS query to the global DNS server to get the answer for the user machine. The user machine gets to know the IP address after receiving the response. 



DNS cache poisoning attack can target local DNS servers. When the local DNS server receives a query which it does not have the corresponding records (first stage), it will inquire the global DNS server (second stage). On receiving the response, the local DNS server will forward the response to the user machine. It also saves this record in its cache (third stage) to avoid inquiring the global DNS again when receiving the same query. However, the DNS server cannot verify the response at the second stage, and this is where the attacker can fool the local DNS server. Pretending as the global DNS server, the attacker can send a spoofed DNS response to the local DNS server with the falsified DNS record. As long as the fake response arrives earlier than the real one, the local DNS server will forward it to the user machine and save the falsified record to its cache. When new queries about the same domain name comes in, the local DNS server will not send a query to the global DNS server again because the corresponding record has been cached. Consequently, it will answer the user machine with the falsified record, until the record expires or the cache is flushed.

For this attack, ten fields, such as \texttt{time to live} values in different layers and resource records, are fuzzed in \texttt{IP} and \texttt{DNS} layers. The testbed contains three machines: a local DNS server whose DNS cache is flushed periodically, a user machine which sends out DNS queries to the local DNS server periodically, and an attacker machine which sniffs for DNS requests sent from the local DNS server and answer them with spoofed responses in the attack scenario, and does nothing in the benign scenario.

In the malicious scenario, we let the user machine to ask for the IP address of one specific domain name from the local DNS server using command $dig$. The domain name is one that does not have a corresponding record on the local DNS server, thus enabling the DNS cache poisoning attack towards it. The attacker machine sniffs for DNS queries with that specific domain name sent out from the local DNS server, and responds them with fuzzed DNS responses with falsified IP addresses. Then the DNS cache gets poisoned and the user machine gets the falsified DNS record. We keep the user machine to send out DNS queries periodically, so that the above process happens lots times and a large amount of data can be generated. However, as discussed earlier, if the local DNS server has the record for the domain name in its cache, it will not send out DNS queries for it, which is why we flush the DNS cache of the local DNS server, so that it remains vulnerable in different iterations. If the attack is successful, the falsified IP addresses can be seen on the results of $dig$.

In the benign scenario, we prepare a list containing 4098 domain names. Then, in each iteration, the user machine randomly chooses one domain name from the list, and ask the local DNS server for its IP address. In order to resemble the malicious scenario, the cache of local DNS server is also flushed periodically so that the local DNS server always needs to communicate with the global DNS server.

The specific domain name used in the malicious scenario and the domain names used in the benign scenario do not overlap. Both the domain names and the IP addresses (falsified or genuine) are excluded during neural network training, so that the neural network will not learn the specific domain name or falsified IP address, which can be treated as signatures by the neural network. Details are provided in subsection~\ref{subsec:DNS-nn}.

\vspace{-4mm}
\subsection{TELNET Session Hijacking Data Generation}
\label{subsec:data-gen_TELNET}
TELNET session hijacking is one of the TCP session hijacking attacks. The most common method of session hijacking is through IP spoofing, in which an attacker manipulates their packets, making them seem to be from the benign client, and inserts commands into an active communication between the benign client and server. This type of attack is possible because authentication typically is only done at the start of a TCP session. One prerequisite for this attack is that the attacker can access the user's token, which enables the attacker to manipulate the network packets. There are several ways of acquiring those tokens, such as packet sniffing, cross-site scripting (XSS), man-in-the-middle (MITM), etc. We have conducted a TELNET session hijacking attack based on packet sniffing in our testbed. Although TELNET may be not commonly used these days, the attack mechanism still holds.

The test bed contains three computers: one Ubuntu machine as the server, one Ubuntu machine as the client, and one Kali Linux as the attacker. There are a total of six fuzzed fields in \texttt{IP} and \texttt{TCP} layers. The server machine is configured to restart the TELNET server periodically, so that any possible problems in the previous session will not affect sessions afterwards.

In the malicious scenario when attacks exist, the client machine is configured to start a TELNET session with the server, execute a command, wait for certain amount of time, then exit and start all over again. Meanwhile, the attacker is sniffing packets sent between the client and the server. When the session is idle, the attacker will try to hijack this session and send a malicious command to the server with fuzzed network packets, trying to start a reverse shell between the server and the attacker. The attacker needs to wait when the session is idle because, for the attack to succeed, packets from the attacker need to arrive at the server machine earlier than those from the client machine. However, malicious packets need to be fuzzed before sending out, so the processing time is potentially longer than that of normal packets. Therefore, to improve the attack success rate, we make the client wait and make the attacker send out packets when the session is idle. The malicious command is to redirect standard input and output over network to the attacker, so that the server takes inputs from and prints outputs to the attacker rather than the client. Therefore, the sign of a successful attack is an established reverse shell.

In the benign scenario with no attacks, we first prepare a list of commands, including file reading/writing, and internet interactions, etc. Before starting the TELNET session, the client randomly picks up three commands from the list, starts a TELNET session with the server, executes those three commands, reads outputs respectively, exits, re-chooses three commands and starts all over again.

One thing worth noting is that, different from other attacks, data collected for TELNET session hijacking attacks always include benign data, even when attacks exist. The reason is that session hijacking require existing sessions to be hijacked. The benign client always exist in the testbed and communicates with the server, otherwise the attacks cannot succeed, let alone collecting data.

\vspace{-2mm}
\section{Detection Models}
\label{sec:detect_with_NN}
In this section, we will discuss the network attacks' detection models trained with our generated data. Specifically, we will discuss how the data is processed before fed to the neural network, the neural network's structure, and training details.

\vspace{-2mm}
\subsection{PtH Detection}
\label{subsec:pth-nn}
To detect PtH attack with neural networks, we have two key insights that help determine the representation of data samples: 1) Network communication for authentication is actually a sequence of network packets in certain order. An earlier packet can affect the packet afterwards. For example, the first several packets between a server and a client may be used to communicate and determine which protocol to use (e.g. SMB or SMB2), and packets afterwards will use the decided protocol. The attack is to get authenticated by the server, which requires a sequence of packets to accomplish. Therefore, each data sample should be a sequence of packets, rather than an individual packet. 2) PtH relies on authentication mechanisms that legitimate users usually don't use. The network packets for the benign and malicious authentication are different. Since both authentication methods use SMB/SMB2 packets, the differences between them thus exist in the fields of the SMB/SMB2 layer. Therefore, data in \texttt{SMB/SMB2} layer is used for PtH detection. In addition, the differences of field values between benign and malicious authentication will be helpful to distinguish them. 
    

\begin{figure}
    \centering
    \includegraphics[width=0.5\textwidth]{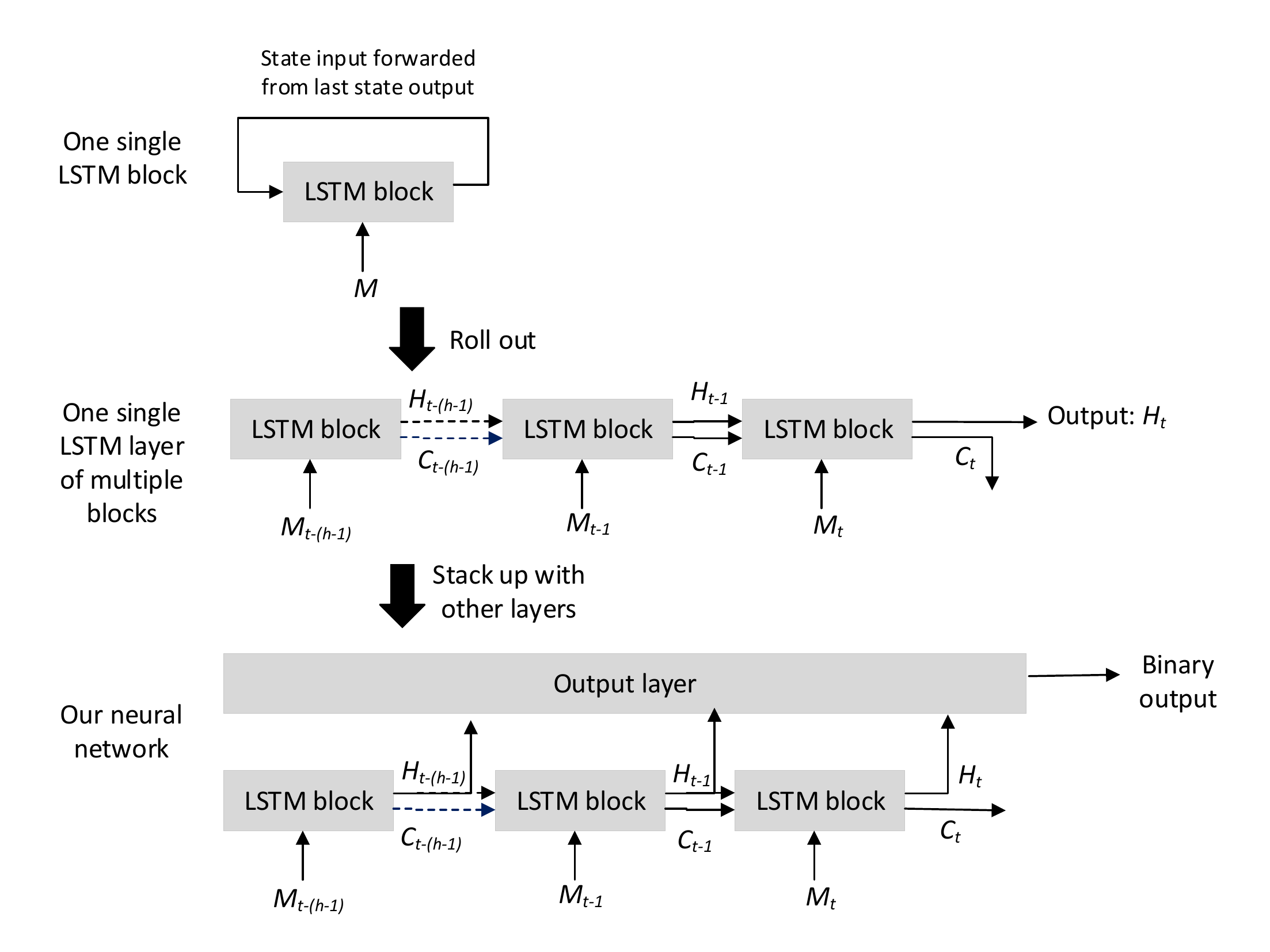}
    \vspace{-6mm}
    \caption{LSTM neural network for PtH detection.}
    \vspace{-4mm}
    \label{fig:classification_NN}
\end{figure}


For this attack, we choose Long-short term memory (LSTM) as the architecture for the neural network. For each packet, we assign a type number to it to represent this packet as a whole. The packet types are defined by fields of interest in the packets. If two packets have the same values in all those fields, then the two packets are given the same packet type number. Otherwise, different numbers are assigned. Please note that we only care about the differences of field values between two packets, but do not care about the values themselves. That's why we are using packet type numbers to represent the packets, rather than examining the content of each packet. The LSTM neural network takes a sequence of network packets' type numbers and outputs the binary label representing whether this sequence is PtH network traffic. Our neural network is presented in Figure~\ref{fig:classification_NN}. The neural network consists of one LSTM layer, which contains multiple LSTM blocks to take inputs, and several fully connected layers, which take inputs from the LSTM layer and produce the final binary output. \textit{M} stands for input; \textit{H} stands for each LSTM block's output; and \textit{C} stands for each LSTM block's state output. Subscripts stand for the time points, in which \textit{h} stands for the window size.


Now that each network packet is represented as a packet type number, the whole network log can be represented as a sequence of packet type numbers. The next step is to create data samples needed for neural network training. We complete this by the following steps: 1) We chop the sequence of packet type numbers to smaller sequences based on the unit of network communications. By identifying the start packet for each benign/malicious network communication, we chop the whole sequence into many \textit{variate-length} sequences, and the beginning of every sequence is a start packet. 2) We further chop the smaller variate-length sequences into one or more \textit{fixed-length} sequences according to the window size and step size. The window size decides how many network packets are represented in one data sample, or, in another word, how many packet type numbers are included. The step size decides the shifting step when chopping large sequence. For example, if a sequence $[1,2,3,4,5,6,7,8]$ is chomped with window size 4, and step size 2, then the resulting small sequences are $[1,2,3,4], [3,4,5,6]$, and $[5,6,7,8]$. 3) Depending on whether the sequence is from a benign traffic or malicious traffic, we assign a \textit{binary label} to it to indicate whether this sequence is benign or malicious. This sequence becomes one data sample for our LSTM neural network. 4) If a fixed-length sequence appear in both the benign and malicious data, this sequence is removed from both of them because it cannot help with the classification. 5) We remove duplicate sequences in both the benign data and malicious data. After these two removal processes, all data samples are finally ready to be used.

A series of neural networks are trained with different data representations (with different windows size and step size) and neural network hyper-parameters. Note that the values of window size and shift step size can affect the number of data samples, so different neural networks may be trained and evaluated with data sets of different sizes. All data sets, before used for training/evaluation, are first balanced by randomly choosing the same amount of benign and malicious data samples.

\vspace{-4mm}
\subsection{ARP Poisoning Detection}
Different from PtH attacks, in which the victim server and attacker/user communicate over a stateful protocol (SMB/SMB2), ARP is a stateless protocol on top of User Datagram Protocol (UDP), and the attack can be accomplished with an individual packet. Therefore, for detecting ARP poisoning attacks, every data sample represents one network packet, rather than a sequence of packets as in PtH detection. In addition, ARP packets have very simple packet structure compared to SMB/SMB2 protocol and it is not clear which fields are critical to detect this attack. For both the malicious and benign data, all the collected ARP packets' length is either 42 bytes or 60 bytes. For the 60-byte packets, the last 18 bytes are all zeros, which means that the meaningful data is only the first 42 bytes. Therefore, we select the first 42 bytes as input data. Every byte is treated as a number when fed to the neural network. As a result, each data sample is an integer list converted from one packet's bytes. The labeling is done towards each data sample, which is the 42 bytes generated from each ARP packet.

To deal with the above mentioned data representation method, fully connected layers are used to construct the neural network. The neural network used for detecting ARP poisoning attack is simple with an input layer, several fully-connected layers, and an output layer. The neural network is trained and evaluated with a balanced data set that has a total of 11984 data samples, of which about 80\% are used as training data, and the rest as test data.

\vspace{-4mm}
\subsection{DNS Cache Poisoning Detection}
\label{subsec:DNS-nn}
Network packets from DNS cache poisoning attack form sessions which consist of queries and answers. Therefore, each data sample should include data from multiple network packets. In addition, it is not clear which fields may be of importance, so we need to investigate the packet content, rather than simply generalizing the packets with packet types as we did in PtH detection. 


Instead of using the whole packet, we use 40 bytes, from byte 15 to 54, of every packet and convert them to integers from 0 to 255 respectively. The first 14 bytes belong to the lowest \texttt{ETH} layer. They are excluded purposefully to rule out the impact of MAC addresses: the MAC address may be treated as signatures to detect malicious packets. The chosen 40 bytes include bytes in \texttt{IP} layer, \texttt{UDP} layer, and part of \texttt{DNS} layer. Only part of \texttt{DNS} layer data is used because the query and records are excluded. Those sections contain the domain names and IP addresses, which are excluded for similar reasons as above: we don't want the neural network to learn the ``malicious'' domain. After the packet data processing, every packet is represented as a fixed-length sequence of 40 integer numbers ranging from 0 to 255. The whole captured network traffic is represented as a sequence of 40-integer sequences. 


The packet processing is different from the process in PtH detection. 1) In PtH detection, each component in a sequence is a packet type number, while in DNS cache poisoning attack detection, each component is one 40-integer sequence. 2) The resulting data sample is also different. In DNS cache poisoning attack, the data sample is a matrix of size $k*40$, which represents $k$ packets and 40 bytes in each packet. Each integer in the matrix ranges from 0 to 255. Therefore, the matrix can also be viewed as a grayscale image of size $k*40$, and each integer at row $i$ column $j$ is the grayscale value of that pixel, corresponding to the $j$th selected byte in the $i$th packet. 

Now that the data samples can be represented as images, we use a convolutional neural network (CNN) to do the classifications, which has been proven to work well in image classification problems. The labeling is done towards each data sample, which is the entire matrix, rather than an individual packet. During the data processing process, the malicious and benign data are processed separately. Matrices generated from malicious data are labeled as malicious, and the similar thing happens when processing benign data.

\begin{figure*}
    \centering
    \includegraphics[width=\textwidth]{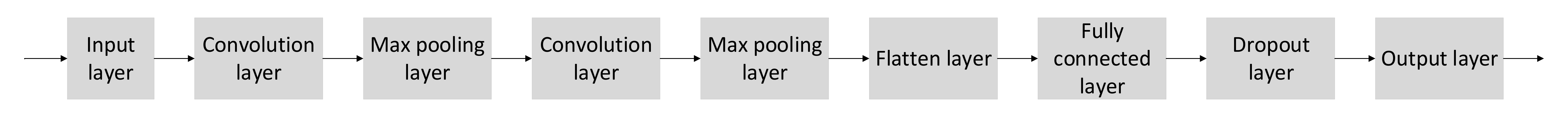}
    \caption{The neural network structure for DNS cache poisoning detection.}
    \label{fig:DNS-NN-structure}
\end{figure*}

Similar to PtH detection, we have trained a series of neural networks with different neural network hyper-parameters and data samples of different window sizes and window steps. That means we can adjust the number of packets $k$ included in each data sample and thus change the size of matrix. The neural network structure is shown in figure~\ref{fig:DNS-NN-structure}.

\vspace{-4mm}
\subsection{TELNET Session Hijacking Detection}
As explained in subsection~\ref{subsec:data-gen_TELNET}, the objective of TELNET session hijacking attacks is to hijack an existing legitimate TELNET session between the server and the client by injecting malicious contents into this session. Therefore, the data collected in the malicious scenario contains commands from both the benign client and the attacker. Based on these observations, we use TELNET packets sent to the server as data samples. TELNET packets sent from the client are labeled as benign data samples, and those sent from the attacker are labeled as malicious data samples.

It is also not clear which fields in the packet may be useful to detect this attack, but some portions of the data should be abandoned: the response TELNET packets sent by the server after receiving command packets, and the TCP packets after the malicious command is executed and the reverse shell is established. Response TELNET packets are abandoned because they are sent by the server passively in response to the command packets from clients or the attacker.
Therefore, they are not necessary if the classification performance is already good with only command packets. As for the TCP packets after the reverse shell is established, strictly speaking, they do not belong to the TELNET session hijacking attack already because the attack has been successfully completed by then. Besides, our way of establishing reverse shells (described in Subsection~\ref{subsec:data-gen_TELNET}) do not use TELNET packets. 
Thus, the only packets marked as malicious are packets containing the malicious command to establish reverse shells. The TELNET packets before these malicious packets, which are TELNET communication between the client and the server, are labeled as benign data samples; The TCP packets after the reverse shell establishment, which are commands from the attacker via the reverse shell, are excluded from the data set.

For every chosen TELNET packet, we exclude data in the application layer and use bytes data in lower layers, such as \texttt{IP} layer and \texttt{TCP} layer. Data in the application layer (\texttt{TELNET} layer in this case) is not used for training because it contains malicious commands. The neural network may be trained to recognize these commands, so we want to rule out their possible impact on the training. 
Furthermore, data in the application layer can sometimes be encrypted. If the attack detection neural network is to be implemented in a router or any types of middleware, it's extremely challenging to make sense of the encrypted data in the application layer. 

For the remaining data in a packet, every byte is treated as a number. Each packet generates a data sample, labeled as benign or malicious, and fed to the neural network. The labeling is easy because the benign and malicious data are not mixed starting from data generation. The neural network is similar to what is used for ARP poisoning detection, with one input layer, several fully-connected layers, and one output layer. The neural network is trained and evaluated with a total of balanced 13098 data samples, of which about 80\% are used as training data samples, and the rest as test data samples.

\section{Evaluations}
\label{sec:eva}
In this section, we will provide evaluation results on our trained models from the previous section. The evaluations contain two parts: 1) evaluations with fuzzed data samples, which are part of the generated dataset; 2) evaluations with data from real network attacks. Note that there are a total of three sets of data used in this paper: training data (fuzzed), test data 1 (fuzzed), and test data 2 (real). They have no overlaps. The training data is used in the previous section for model training. Here in this section, only test data 1 and 2 are used for evaluations.

Our evaluation metrics include the confusion matrix, accuracy, F1 score, and false positive rate. Confusion matrix provides a direct view of how many positive/negative data samples are correctly/incorrectly classified. Accuracy and F1 score are two of the most used metrics for evaluating the performance of neural networks. F1 score is especially useful when the data set is not balanced, as is the nature of real-world network traffics. It is calculated from the precision and recall of the test, where the precision is the number of correctly identified positive results divided by the number of all positive results, including those not identified correctly, and the recall is the number of correctly identified positive results divided by the number of all samples that should have been identified as positive. False positive rate is critical in real-world implementations because enterprises have high motivations to avoid false positives so that normal operations will not be interrupted by mistake.

\subsection{Evaluation on fuzzed data}

\begin{table}
    \centering
    \caption{Confusion matrix of PtH detection results}
    \begin{tabular}{lll}
        \hline
                           & Predicted   negative & Predicted   positive \\ \hline
        Condition negative & 289                  & 3                    \\
        Condition positive & 2                    & 345                  \\ \hline
    \end{tabular}
    \label{tab:con-mat-pth}
\end{table}

For \textbf{PtH} detection, the best-performing neural network is trained by a training set of 1917 data samples and evaluated by a test set of 639 data samples. Table~\ref{tab:con-mat-pth} is the confusion matrix of the detection results on the test set. Condition negative and positive represent labels from the ground truth, predicted negative and positive represent labels of the model output. The accuracy is 99.22\%, the F1 score is 0.9916, and the false positive rate is 0.86\%. The major reason for misclassifications is successive write requests, which is when the payload is being uploaded in the malicious scenario, or when some documents are being uploaded in the benign scenario. In a word, our trained model is not very effective when distinguishing benign or malicious file upload.

\begin{table}
    \centering
    \caption{Confusion matrix of ARP poisoning detection results}
    \vspace{-4mm}
    \begin{tabular}{lll}
    \hline
                       & Predicted   negative & Predicted   positive \\ \hline
    Condition negative & 1188                 & 5                    \\
    Condition positive & 1                    & 1206                 \\ \hline
    \end{tabular}
    \label{tab:con-mat-arp}
\end{table}

For \textbf{ARP poisoning} attacks, Table~\ref{tab:con-mat-arp} is the confusion matrix of the detection results on the test set. The accuracy is 99.75\%, the F1 score is 0.9975, and the false positive rate is 0.41\%. The misclassified six packets include two ARP queries and four ARP responses, one of which is a broadcast packet. The broadcast ARP response is the false negative, and the rest are false positives.

\begin{table}
    \centering
    \caption{Confusion matrix of DNS cache poisoning detection results}
        \vspace{-4mm}
    \begin{tabular}{lll}
    \hline
                       & Predicted   negative & Predicted   positive \\ \hline
    Condition negative & 3878                 & 11                    \\
    Condition positive & 10                    & 3833                 \\ \hline
    \end{tabular}
    \label{tab:con-mat-dns}
\end{table}

For \textbf{DNS cache poisoning} attack detection, the best-performing neural network is trained and evaluated with a total of 38660 data samples. Table~\ref{tab:con-mat-dns} is the confusion matrix of the detection results on the test set. The accuracy is 99.73\%, the F1 score is 0.9973, and the false positive rate is 0.28\%. Misclassifications majorly consist of packets with the set $Fin$ flag in $TCP$ layer.

\begin{table}
    \centering
    \caption{Confusion matrix of TELNET session hijacking detection results}
    \vspace{-4mm}
    \begin{tabular}{lll}
    \hline
                       & Predicted   negative & Predicted   positive \\ \hline
    Condition negative & 1296                 & 1                    \\
    Condition positive & 1                    & 1324                 \\ \hline
    \end{tabular}
    \label{tab:con-mat-telnet}
\end{table}

For \textbf{TELNET session hijacking} detection, Table~\ref{tab:con-mat-telnet} is the confusion matrix of the detection results on the test set. The accuracy is 99.92\%, the F1 score is 0.9992, and the false positive rate is 0.08\%. The misclassification only happens to packets sent from the DHCP server to the user machine, which are beyond our original intention of detecting malicious packets sent to the local DNS server. We also did one additional experiment by including the response TELNET packets from the server. The data processing methods, neural network structure, training hyper-parameters, etc. are the same. The results show minor improvement over the case without response TELNET packets: the accuracy is 99.96\%, the F1 score is 0.9996, and the false positive rate is 0\%.

\subsection{Evaluation on real attack data}
To launch real \textbf{PtH} attacks, we leverage the PtH script with reverse TCP payload, and the attack is successfully launched for 192 times. That is, each of these attacks successfully establishes a reverse shell. The total number of captured network packets is 66903, of which 14512 are SMB/SMB2 packets, and 2742 data samples are generated with respect to the parameters of the best-performing model. As described earlier, in one PtH session, there exist hundreds of network packets, which may result in multiple data samples. Because we do not know the threshold to classify a session as benign or malicious, we tried a series of thresholds: 30\%, 40\%, 50\%, 60\%, and 70\%. That is, in one session, if 30\%, 40\%, 50\%, 60\%, or 70\% of the data samples are classified as malicious, then we determine that this session is classified as malicious, and thus the PtH attack is successfully detected. The results show that, when the threshold is set to 30\%, 40\% or 50\%, all (100\% detection rate) of the 192 PtH attacks are successfully detected by the trained neural network; when the threshold is set to 60\%, 189 (98\% detection rate) out of 192 are successfully detected; when the threshold is set to 70\%, 180 (93\% detection rate) of 192 are successfully detected. 

To evaluate the trained model against real \textbf{ARP poisoning} attack, we launch the ARP poisoning attack for 10000 times using \textit{netwox}, a common tool in Linux which can be used to continuously send out spoofed ARP responses. A total of 65763 relevant ARP network packets are captured. Different from PtH, ARP poisoning can be achieved with a single packet, so the detection is also based on single packets. The results show that 8748 (87\% detection rate) of 10000 attacks are successfully detected. The performance is lower compared to that of the fuzzed data. One possible reason is that the real data is collected from a different LAN, with different DHCP configuration with different valid IP and MAC addresses. If the neural network is trained and deployed on the same LAN, the evaluation results on fuzzed and real data may be closer.

To evaluate the trained model against real \textbf{DNS cache poisoning} attack, we launch the DNS cache poisoning attack for 223 times using \textit{netwox}, which can also be used to sniff for DNS queries and send out spoofed DNS responses. The total number of captured DNS packets are 2489, and the results show that all (100\% detection rate) of the 223 attacks are successfully detected. 

To evaluate the trained model against real \textbf{TELNET session hijacking} attack, we launch the attack for 242 times using \textit{netwox} again. An existing TELNET session is the prerequisite for TELNET session hijacking attack, but the attack itself can be achieved with only one packet. Therefore, during one TELNET session hijacking, if at least one packet is classified as malicious, then the whole session is classified as malicious. The total number of captured TELNET packets is 3829, of which 308 are malicious TELNET packets. Out of those 308 malicious packets, 307 are correctly classified as malicious, corresponding to 241 malicious sessions. In short, out of 242 successful TELNET session hijacking attacks, 241 (99.5\% detection rate) are successfully detected by our trained neural network. It is not clear why that one packet evades the detection, though. On one hand, those 308 packets only differ from each other in fields $source port$, $sequence number$, $acknowledge number$ and $checksum$ of the $TCP$ layer. On the other hand, those fields themselves do not seem to hold any specific meanings.

\vspace{-2mm}
\subsection{Evaluation with Support Vector Machines}
Though we apply protocol fuzzing to generate data originally for deep learning, the data can also be used for training traditional machine learning models. For demonstration, in this subsection, we will show evaluation results on real attack data of support vector machines (SVMs) trained with our generated data. A series of SVM models with different training parameters are trained. We select the best-performing models with widely-used procedures, and the evaluation results are presented in Table~\ref{tab:eva-svm-real}. From Table~\ref{tab:eva-svm-real}, we can get the following observations: 1) For PtH, DNS cache poisoning, and TELNET session hijacking detection, SVM model seems to be substantially worse than neural network; 2) for ARP poisoning detection, SVM model is about 3\% better than neural network.

\begin{table}[]
    \centering
    \caption{SVMs' evaluation results on real attack data.}
    \begin{tabular}{lll}
        \hline
        Network Attacks          & \begin{tabular}[c]{@{}l@{}}Total number\\ of attacks\end{tabular} & \begin{tabular}[c]{@{}l@{}}Number of\\ detected attacks\end{tabular} \\ \hline
        PtH                      & 192                                                               & 136 \footnote{The threshold is set to 50\%.}                                                                  \\
        ARP poisoning            & 10000                                                             & 9028                                                                 \\
        DNS cache poisoning      & 223                                                               & 201                                                                  \\
        TELNET session hijacking &    242                                                            & 235                                                                  \\ \hline
        \end{tabular}
        \label{tab:eva-svm-real}
\end{table}

\section{Interpreting Why the Detection Models Achieved High Detection Accuracy}
\label{sec:interpretation}

The results presented in the previous section show that the trained (network attack detection) models have achieved high detection accuracy, as measured by metrics such as 
accuracy score, F1 score, and false positive rate. 
In this section, we would like to provide a few domain-specific interpretations on why the detection models could achieve high detection accuracy. 
In particular, we seek to answer two research questions: (1) Whether fuzzed packet fields indeed influence model accuracy? (2) Whether the protocol fuzzing approach has the ability to generate comprehensive data?    



\subsection{Influence of Fuzzed Packet Fields on Detection  Accuracy}
\label{subsec:influential}

One of the problems we are curious about is how influential the fuzzed fields are to the accuracy of the trained neural networks. To answer this question, we have leveraged {\em permutation feature importance} \cite{breiman2001random}, a widely-adopted procedure to make machine learning more interpretable, to measure the increase or decrease in the accuracy of the model after we permute a particular feature's values. We obtained the permutation feature importance 
measurements for ARP poisoning and TELNET session hijacking identification. For DNS cache poisoning, 
we visualized the learned features in the first convolutional layer, which provide insights about which pixels of the ``image'' are emphasized.

Permutation feature importance is a model inspection technique that can be used for fitted models. It is defined as the decrease in a model score, which is the F1 score in our experiments, when a single feature value is randomly shuffled~\cite{breiman2001random}. This technique benefits from being model agnostic and can be calculated many times with different permutations of the feature, and is often used to inspect how influential some features are to the model in various deep learning applications~\cite{galkin2018human,liu2020deep,zamani2019pm2}. The drawback is that, when several features are correlated, the measurement becomes inaccurate. Our results show that, for ARP poisoning, the top three most influential features (bytes) all belong to fuzzed fields, which are \texttt{destination MAC address}, \texttt{target MAC address}, and \texttt{target IP address}; for TELNET session hijacking, one of the three most influential features is fuzzed, which are \texttt{timestamp value}, \texttt{timestamp echo reply}, and \texttt{urgent pointer}, and \texttt{urgent pointer} is fuzzed.

In summary, our experiments show that the 3 most influential features for each of the trained models usually involve one or more fuzzed packet fields. However, as for the CNN for DNS cache poisoning identification, the fuzzed field, \texttt{time to live} value in the \texttt{IP} layer, is one of the fields that are least influential.
Although our experiments show no fuzzed field is most influential on the first layer (of the model) in detecting DNS cache poisoning, the possibility that a fuzzed field is most influential on a deeper layer is not denied either. 
Deep learning models are still mostly ``black boxes''. Furthermore, not all fuzzed fields may be presented in the data samples. For example, for PtH, the fields used in the mapping (and embedding) do not concern any fuzzed fields.
Another possibility is that, though some fuzzed fields' influences are not directly reflected from our measurements above, the fuzzed fields may affect other fields, which are in turn influential to the models. 

\subsection{Can Protocol Fuzzing Generate Comprehensive Training Data?}

A comprehensive training data set is a key factor in many 
successful applications (e.g., image classification) of deep learning. 
However, how to formally assess the comprehensiveness of the training data in deep-learning-based detection of network attacks is an issue largely 
neglected in the literature. 
In this subsection, we seek to take the first step towards 
systematically addressing this issue. 
In particular, we introduce a new notion, called \textbf{``covered-by"}, 
to capture an inherent connection between protocol fuzzing and 
training-data comprehensiveness.

The intuition behind this notion is as follows. 
First, 
a network packet is essentially a set of fields. 
Although real-world packets are usually not byte-to-byte identical to 
the packets generated through protocol fuzzing, the content in a particular field of a real-world packet may be identical to 
the corresponding field of a fuzzed packet. 
Hence, it seems appropriate to do field-content-based data 
comprehensiveness assessment. 
Second, one major advantage of deep learning 
is that it requires less feature engineering than conventional 
machine learning techniques (e.g., Support Vector Machines). 
Accordingly, although a data sample in the training set 
involves the information from multiple fields, the deep learning
algorithm typically treats each element (e.g., a byte or a pixel) in 
the data sample in 
a unified way, and does not pre-extract semantic-carrying features. 
Based on this observation, we do not assess data comprehensiveness
based on a space of semantic-carrying features. 

Based on the above intuition, we define the 
notion of ``covered-by" as follows: 1) When a real-world network packet is being ``examined" 
by a deep learning model, it will need to be firstly transformed 
to a data sample $a$ (or a portion of $a$) expected by the model. 
An element in the data sample is {\em covered-by} 
a data sample $b$ generated through the proposed protocol fuzzing method
if A) the element's content is identical to the corresponding
element in $b$, B) the 
packet field used to derive 
the element in $a$ has the same content as the field used to derive the corresponding element in $b$, 
and C) the above-mentioned packet field is a being-fuzzed
field in the proposed protocol fuzzing.  
Note that since the definition 
intends to capture an inherent connection between protocol fuzzing
and data comprehensiveness, it ignores the elements that are 
not derived from a being-fuzzed field.  
2) Given a set of data samples $S_1$ (generated through real-world network packets) and a training set of data samples $S_2$ generated through 
the proposed protocol fuzzing method, the {\em coverage rate} is defined as the ratio of $x$ to $(x+y)$. 
Here $x$ is the total number of the elements in $S_1$ 
that are covered-by one or more data samples in $S_2$. 
And $y$ is the total number of the elements in 
$S_1$ that are derived from a being-fuzzed field 
but not covered-by any data sample in $S_2$. 

It should be noticed that coverage rate is not applicable to 
the deep learning algorithms that employ a mapping from an entire network packet to 
a type assigned with a particular type number. 
Accordingly, because such a mapping is done
when detecting PtH, we should not use 
coverage rate to 
measure the comprehensiveness of the training data 
generated for detecting PtH.

{\bf Measuring the coverage rates. } 
To measure the coverage rates for the four network attacks, we have collected additional non-fuzzed data and measured the coverage rates with respect to the fuzzed data collected in Section~\ref{sec:data-gen}. The non-fuzzed data are collected by launching each kind of network attack with different parameters (e.g., IP addresses, MAC addresses, process names, etc.) for hundreds of times so that data collected in different attack sessions are not identical to each other.

We have measured the coverage rates for ARP poisoning, DNS cache poisoning, and TELNET session hijacking, respectively. 
Our results show that, for all the three network attacks, the coverage rates are all 100\%. 
In addition, we also assess the comprehensiveness of the training data generated for detecting PtH in 
an ad hoc way, and found that for each being-fuzzed field, 
every value observed in the non-fuzzed network packets
appeared at least once in the fuzzed data. 
In summary, our results show that a correlation between high detection accuracy and high coverage rates exists.

\vspace{-2mm}
\section{Discussions and Limitations}
\label{sec:dis-and-lim}
In previous sections, we have shown that our new approach, protocol fuzzing, is helpful in generating data on which neural networks can be trained. However, our approach still has some limitations.

\textbf{Efficiency: }Training a neural network requires a large amount of data samples. However, the number of data samples can be affected in many ways. For example, protocol fuzzing in nature cannot guarantee that all malicious/benign activities are successful. Although the fuzzed values are in a valid range, the network packets with fuzzed values may still get rejected by the server or trigger some unexpected circumstances, leading to an interrupted session. Those data are probably useless as discussed in section~\ref{subsec:data-gen_PtH}. Also, the removal of duplicate (same data in one class) and double-dipping (same data among different classes) data samples will also affect the number of data samples. In a word, not all collected data can be used as data sample for neural network training. 

Another factor that affects the efficiency is the time consumed by each benign/malicious activity. Except for some simple activities like MAC-IP address resolving with only several ARP packets, other complicated activities need time to get carried out, especially those containing hundreds or more network packets. Also, depending on the mechanism of packet processing, the client/server may also need more time before it can respond. For example, in PtH data generation, one successful attack contains 300 to 400 packets (and not all of them are usable to generate data sample), and some time intervals between adjacent packets can be as large as 0.5 second. In addition, in our experiments, we manually inserted idle time intervals. The purpose is to stabilize the activity. For example, in PtH experiment, after starting the exploitation, the foreground is put to sleep for 25 seconds. This time interval is reserved so that the exploitation can continue to run to reach a successful end. If this time interval is removed or too short, then the attack process is very likely to end at the middle of exploitation. In a word, each data generation iteration takes time to complete.

Because of the two points above, our data generating efficiency is not very high. Take PtH as an example. We spent about 4 days running 5,000 attack iterations, of which 611 failed. The total amount of network packets captured is 497,956, of which 103,718 are SMB/SMB2 packets that may be used to generate data sample. However, the final number of data samples is just 2,556.

\textbf{Neural networks for various network attacks: }Though we have verified our idea on four chosen network attacks, we trained separate neural networks for different attacks and did not train a neural network to detect various network attacks. It is difficult to train such a neural network because different network attacks have different characteristics, which may need different data representations and neural network architectures, as shown in section~\ref{sec:detect_with_NN}. 

\textbf{Interpretability: }In subsection~\ref{subsec:influential}, we have tried to interpret the trained neural networks in two widely-used ways. However, there are still other interpretation techniques we have not tried, which may show some further insights about the neural networks and the attacks themselves. Our future work may provide additional interpretations about the trained neural networks.

\vspace{-2mm}
\section{Conclusion}
\label{sec:conclusion}
In this paper, we first discussed detection of which type of network attack is suitable for deep learning, and demonstrated it with four specific network attacks of that type. We have trained neural networks with data generated using protocol fuzzing and also evaluated the data and the trained model. We also present interpretation of our models, discussions and limitations of our experiments and approach. Our results show that the proposed new approach, protocol fuzzing, can generate high-quality data on which deep learning models can be trained to detect some network attacks.

\section*{Disclaimer}
This paper is not subject to copyright in the United States. Commercial products are identified in order to adequately specify certain procedures. In no case does such identification imply recommendation or endorsement by the National Institute of Standards and Technology, nor does it imply that the identified products are necessarily the best available for the purpose.



\end{document}